\documentclass[aps,twocolumn,amsmath,amssymb,pre,floatfix]{revtex4-1}
\usepackage{amssymb}
\usepackage{amsmath}
\usepackage{graphicx}
\usepackage{epsfig}
\usepackage[utf8]{inputenc}
\usepackage{dcolumn}
\usepackage{bm}
\usepackage{colortbl}
\begin{document}

\title{An Atomistic First-Principles Density Functional Theory Model for Single Layer Dry \textit{Stratum Corneum}}

\author{Erika T. Sato, Neila Machado, Daniele R. Araújo, Luciana C. Paulino and Herculano Martinho}

\email{herculano.martinho@ufabc.edu.br}

\affiliation{Centro de Ciências Naturais e Humanas, Universidade Federal do ABC, Av. dos Estados 5001, Santo André-SP, 09210-580, Brazil}

\begin{abstract}

Many questions concerning the biophysical and physiological properties of skin are still open. Skin aging, permeability, dermal absorption, hydration and drug transdermal delivery, are few examples of processes with its underlying mechanisms unveiled. In this work we present a first-principles density functional quantum atomistic model for single layer stratum corneum (SC) in order to contribute to unveil the molecular interactions behind the skin properties at this scale. The molecular structure of SC was modeled by an archetype of its hygroscopic proteic portion inside of the corneocytes, the natural moisturizing factor (NMF), coupled to glycerol molecules which represent the lipid fraction of SC. The vibrational spectra was calculated and compared to Fourier-Transform Infrared Absorption spectroscopy (FTIR) experimental data obtained on animal model of SC. We noticed that bands in the fingerprint region (800-1800 cm$^{-1}$) were correctly assigned. Moreover, our calculations revealed the existence of two coupled vibration between hydroxyl group of lipid and NMF methylene (1120 and 1160 cm$^{-1}$), which are of special interest since they probe the lipid-amino acid coupling. The model was also able to predict the shear modulus of dry SC in excellent agreement with the reported value on literature. Others physical/chemical properties could be calculated exploring the chemical accuracy and molecular resolution of our model. Research in dermatology, cosmetology, and biomedical engineering in the specific topics of drug delivery and/or mechanical properties of skin are examples of fields that would potentially take advantage of our approach.


\end{abstract}

\maketitle

\section{Introduction}

Skin is a complex, integrated, and dynamic organ that has many functions that go far beyond its role as a barrier to the environment\cite{hongbo}. It protects the body against external chemical and physical factors, takes part in the metabolic processes, plays a resorptive and thermoregulatory function, and it partakes in immunological processes\cite{hongbo,boer2016structural}. The basic structure of the skin comprises three component layers: epidermis, dermis, and subcutis or hypodermis\cite{Burkitt2014}. The epidermis is the outer layer of the three layers that make up the skin\cite{Burkitt2014}. The epidermis is divided into four layers, starting at the dermal junction with the basal cell layer and eventuating at the outer surface in the stratum corneum (SC)\cite{JamesG.MarksJr.2013}.

A large number of research tools is available for the investigation of biophysical parameters of skin, e.g., transepidermal water loss, electrical impedance, Raman spectroscopy, confocal spectroscopy, optical coherence tomography, magnetic resonance imaging, among others\cite{byrne2010bioengineering}. In spite of large experimental data available, several issues such as permeability and dermal absorption processes, remain unsolved. Moreover, skin aging processes, hydration and the role of water in life as well still lack clarity\cite{ball2008water}. 

The general morphology of SC is described within the “bricks and mortar” model (see Fig.1a and ref. \cite{del2011clinical}). The corneocytes represent the bricks and the intercellular lamellar lipid membrane represents the mortar between the bricks\cite{del2011clinical}. Corneocytes are primarily proteic fiber comprised of keratin macrofibrils. They are protected externally by a cornified cell envelope, and are cohesively held together by corneodesmosomes\cite{del2011clinical}. Ceramides (40 – 50 \%), cholesterol (25 \%), and fatty acids (10 – 15 \%) comprises the main components of the intercellular lamellar lipid membrane\cite{del2011clinical}. SC is composed by ~ 20\% of water, of which a fraction is tightly bound to hygroscopic molecules (the natural moisturizing factor, NMF) and lipids in the skin. NMF compounds are present in high concentrations within corneocytes and represent up to 20\% to 30\% of the dry weight of the SC. Much of the NMF is represented by amino acids (40\%) and their derivatives derived from the proteolysis of epidermal filaggrin\cite{Verdier-Sevrain2007}. Approximately 34\% of the amino acids are neutral\cite{Verdier-Sevrain2007}. The role of the NMF lies in the fact that its constituent chemicals are highly water soluble and hygroscopic. It plasticizes the SC keeping it resilient and preventing cracking and flaking due to mechanical stresses and is a hydrolitic medium for several important enzymatic reactions in the skin\cite{del2011clinical,Verdier-Sevrain2007,Jokura1995}.

\begin{figure*}[th!]
\centering
	\includegraphics[width=12.0cm]{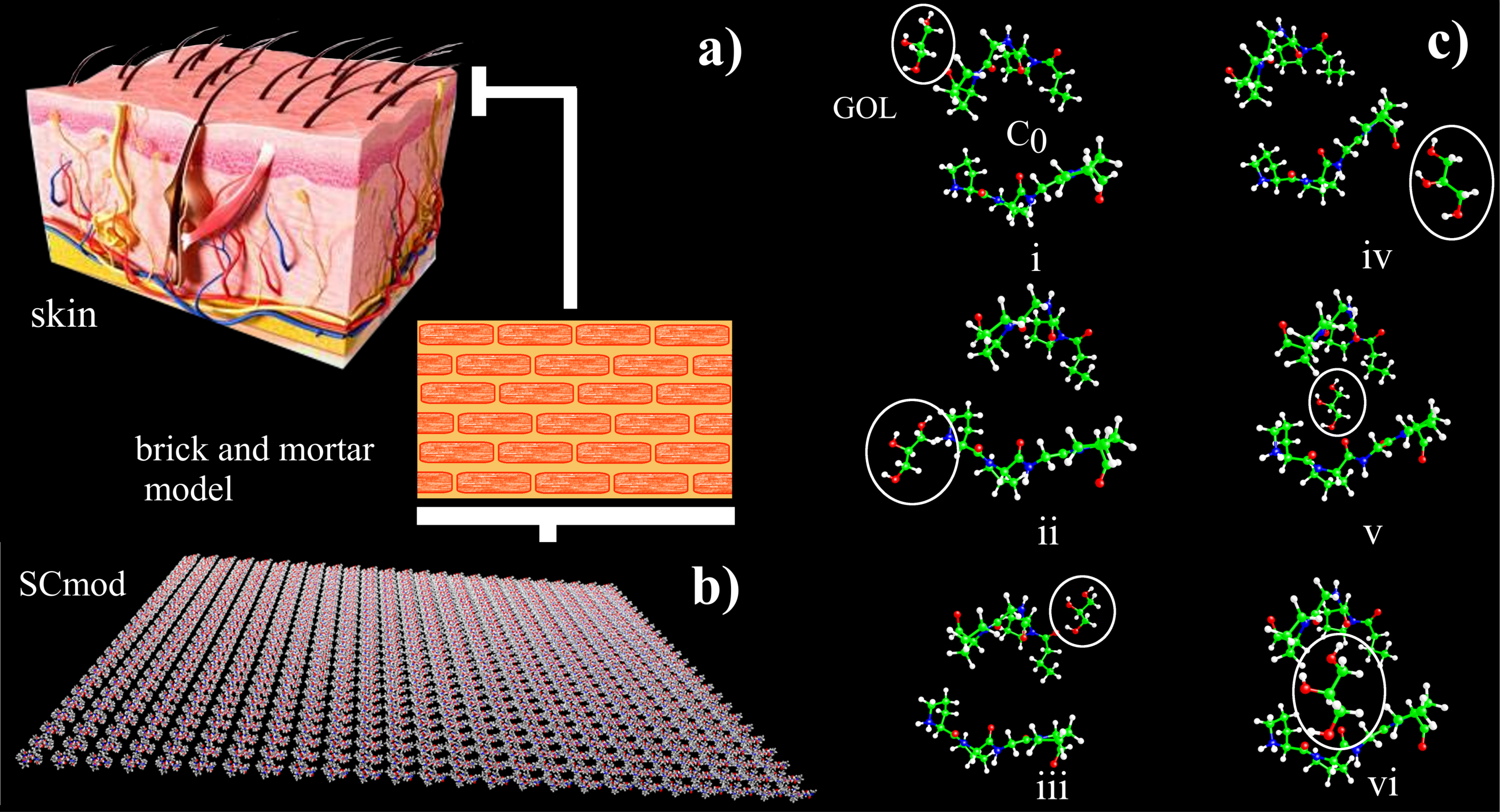}
	\caption{a) Skin representation (source: https://www.myvmc.com/anatomy/human-skin/) including the stratum corneum (SC) as bricks and mortar model. b) Computational DFT-based SC model (SCmod), which comprises a two-dimensional reticulate of unit cells subjected to periodic boundary conditions. c) SCmod unit cells tested in the present work. A glycerol (GOL) molecule was added to previously published C0 model (see ref.\cite{sato}) at 6 positions (indicated in i) to vi) .}\label{skin}
\end{figure*}

Histological changes in the subcutaneous structures which give rise to biomechanical alterations are reported to occur along the aging process. The tissue roughness and elasticity are the main biomechanical parameters employed to determine the aging degree\cite{Maiti2016}. The above-mentioned changes are correlated to collagen site properties since collagen is directly related to structural functions of skin. In epidermis it was reported decreasing synthesis of keratinocytes, filaggrin, tranglutaminase, and collagen type VII due to aging\cite{Maiti2016}. The collagen VII is also crucial for the function and stability of the extracellular matrix as it is an anchoring fibril collagen, where it binds type I and type III collagen, providing stability to the interstitial membrane and the basement membrane extracellular matrix structures\cite{Karsdal2016}. The basal layer is responsible in large extension by the skin mechanical properties\cite{Karsdal2016}. Amao \textit{et al.}\cite{Amano2009} relates that the basal layer of sun-exposed skin becomes damaged, multilayered, and partly disrupted compared with that of sun-protected skin. Basal layer plays important roles in maintaining a healthy epidermis and dermis, and repeated damage destabilizes the skin, accelerating the aging process\cite{Amano2009}.

Skin is the the most accessible organ for drug delivery purposes. The transdermal delivery could for example, reduce first-pass metabolism associated with oral delivery, and is not painful as injections\cite{schoellhammer2014skin}. However, the major limitation for active compounds penetration through the skin is overcoming the SC. It serves as a rate-limiting lipophilic barrier against the uptake of chemical and biological toxins, as well as trans-epidermal water loss\cite{fox2011transdermal,herman2015essential}.

A number of factors affects the dermal absorption, including, e.g., type of skin; physical-chemical properties of delivery systems; skin moisture level; external temperature; skin pre-treatment; among others. Thus, numerous in-vitro and in-vivo models are used to examine the penetration of active compounds through the skin\cite{herman2015essential,godin2007transdermal}. 
	
In-silico approaches are in principle able to predict adsorption, distribution, metabolism, excretion, and toxicity on the basis of input data describing physical-chemical properties of the compounds to be delivered and on physiological properties of the exposed skin\cite{Baba2015,Selzer2015,Hatanaka2015,Verheyen2017}. In-silico models also represent promising strategies to reduce or to avoid the usage of animal studies for drug development\cite{frohlich2016alternatives}.
	
Classical molecular dynamics have been used to perform permeability studies at molecular level\cite{machado2016assessment,rocco2017molecular}. However, these approaches can obscure some relevant physical-chemical data from the environment since in the hard-sphere approximation the neast-nearest interactions are not implicitly considered. For example, the explicit structure of water and its interactions with the neighboring are not accounted for\cite{sato}. At scale of real biology they are in fact only a small part of the overall picture\cite{saunders2013coarse}. The main drawbacks of the these computational methods are the extensive computational burden\cite{MC1}, absence of realistic phase function, and elastic light scattering models\cite{MC2}.
	
Atomistic models based on quantum mechanics calculations have the greatest predictive capability for materials properties. Detailed atomistic models for biological tissues are rare in the literature due to their inherent complexity (aperiodicity and large number of atoms). In order to correctly describe soft tissues and be computationally viable the atomistic model should be small enough ($<$ 150 atoms for actual quantum methods and computational processing capabilities) and include water-water and water-molecule effects\cite{sato}. From the quantum point of view, the Density Functional Theory (DFT) has been shown to be a promising method. It is considered a mature technique with good experimental reproducibility\cite{cole}. In biological applications, their results are also very accurate. DFT methods have been employed to unravel complex subjects such as biomolecular structure and electronic properties, and the development of medicinal chemistry. Furthermore, studies of nanomaterials at the biomolecular interface have been considered new challenges and opportunities for application of DFT methods\cite{cui}.
	
Previously\cite{sato,Sato2018} we presented a DFT-based atomistic computational model for dermis tissue (STmod), which was able to successfully explain important experimental structural and general biochemical trends of normal and inflammatory oral tissues. Such model was pioneer in including quantum details for macroscopic tissue modeling. In the present work we report a SC model (SCmod) built from the STmod as starting point. The stability of some proposed variations in SCmod was also considered and compared to FTIR experimental data.
	
\section{Methodology}

\subsection{Computational details}

It is computationally inviable represent the full molecular complexity of SC with available DFT methods. Our choice was to represent the SC by its hygroscopic proteic portion inside of the corneocytes, the NMF. The choice of a suitable unit cell, which forms an infinite reticulate representing SC (see Fig. 1b), was the main step for building the model. We considered the proline-rich C$_0$ and D$_0$ unit cells of STmod \cite{sato} as starting point. Proline is a higroscopic neutral amino acid. It is present in NMF and collagen structures\cite{Burkitt2014}. The C structure has orthorhombic symmetry (P$2_{1}2_{1}2_{1}$ space group); unit cell parameters $a = 13.5$ \AA, $b = 11.5$ \AA, and $c = 10.5$ \AA, and internal cage volume (oblate spheroid) $V_{c} = 1.857.3$ \AA$^{3}$. Another structural variant is the compact version of C structure, named D model, with unit cell parameters $a = 17.4$ \AA, $b = 13.3$ \AA, $c = 9.8$ \AA and $V_{c} =182.8$ \AA$^{3}$.

The inclusion of one glycerol (GOL) molecule in the unit cell mimetics the lipidic SC structure. We notice that the inclusion of a larger number of GOL molecules become computational cost of the simulation prohibitive. GOL molecule was added at 6 possible positions (indicated in i) to vi) in Fig. 1c). The periodic boundary conditions were established in two-dimensions (see Fig.1b). Thus, our model represents a single layer of SC. Each structure was first optimized using molecular mechanics. The MMFF94s force field\cite{halgren1999mmff} uses the van der Waals term and includes cubic and quartic terms in the bond stretch, and cubic terms in angle bending potential energy. MMFF94s produces relatively flat dynamically averaged structures\cite{hanwell2012avogadro}. The models were pre-optimized in a first step using molecular mechanics with the MMFF94s force field implemented in the Avogadro software\cite{halgren1999mmff}.

DFT\cite{hohenberg1964inhomogeneous} was used in order to obtain equilibrium geometries and harmonic frequencies. Calculations were implemented in the CPMD program\cite{cpmd} using the BLYP functiona \cite{lee1988development} augmented with dispersion corrections for the proper description of van der Waals interactions\cite{lee1988development,von2005performance}. For all simulations, the cutoff energy was considered up to 100 Ry. The linear response for the values of polarization and polar tensors of each atom in the system was calculated to evaluate the eigenvectors of each vibrational mode. In order to verify the feasibility of the model, vibrational calculations were performed and were compared to Fourier-Transform Infrared Absorption Spectroscopy (FTIR) experimental data obtained from SC animal model.
	
\subsection{Animal model for skin}

Full thickness pig ear skin was obtained from a local slaughter-house, and protocol was approved by Animal Ethical Committee (\#007/2011). The SC was isolated from the dermis according to the methodology previously described by Kaushik and Michniak-Kohn\cite{Kaushik2012} with minor modifications. The skin was excised post-sacrifice from the outer part of pig ears separated from the underlying cartilage with a scalpel and skin samples were kept in a water bath at 37 $^{0}$C with 5 mM Hepes buffer and 154 mM NaCl, pH 7.4, supplemented with trypsin (0.1\%) for at least 4 hours. Subsequently, the skin segments were washed with the buffer solution (without trypsin) for 1 h. The skin samples were dispersed in deionized water, dried on filter paper and stored under silica atmosphere until use. The layers obtained in this method are $\sim$ 100 $\mu$m thick enclosing $\sim$ 80 $\mu$m of epidermis and $\sim$ 20 $\mu$m of SC\cite{Jacobi2007}.

\subsection{FTIR}

The spectra were collected in the Varian 610 FT-IR Micro-spectrometer in the reflectance method with spectral resolution of 2 cm$^{-1}$, 200 scan per sample and 800 scans per background. The sample holder was a platinum recovered surface plate. Three spectra per sample were obtained, and each one was baseline corrected manually with Fityk software\cite{fityk} and normalized to 1. Then the average spectrum was calculated. For data comparison with computational calculations, each spectrum was simulated as a convolution of Gaussian lineshape peaks centered on the calculated frequencies using Fityk software. The linewidth was chosen to be 20 cm$^{-1}$.

\section{Results and discussion}

In order to infer about the influence of the GOL position on the unit cell stability we calculated the GOL binding energy ($E_B$) according to 

\begin{equation}
E_{B}=E_{STmod}-(E+E_{GOL})
\end{equation}

where $E_{STmod}$ is the energy of the SCmod structure (using C or D STmod unit cells), $E$ is the $C_0$ or $D_0$ energy, and $E_{GOL}$ is the energy of glycerol molecule. Results are shown in Figure 2. All energies appeared to be negative, which indicates that no external work needs be done to introduce one GOL inside the amino acid cage since and the GOL molecule spontaneously tends to bond in these sites. Thus, these structures are viable to accommodate glycerol molecules. 

\begin{figure}[th!]
\centering
	\includegraphics[width=8.0cm]{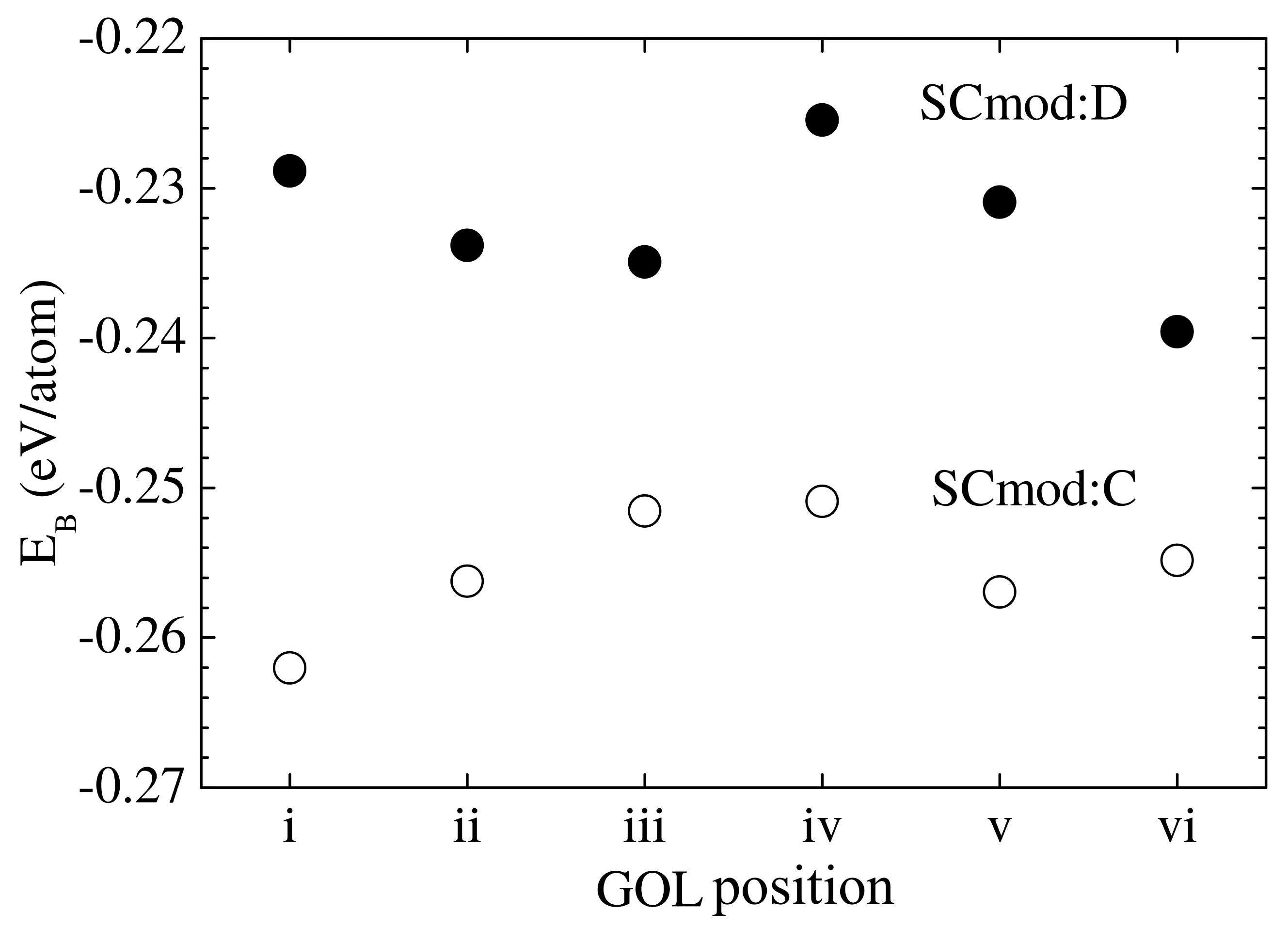}
	\caption{Energy binding $E_B$ (eq. 1) as function of the glycerol (GOL) position for stratum corneum SCmod. The positions are indicated on Fig. 1c). As described in the section 2.1, $C_0$ and $D_0$ STmod unit cells options (ref. \cite{Maiti2016}) were used as starting point for SCmod. They are indicated as SCmod:C and SCmod:D, respectively.}\label{fig2}
\end{figure}

However, since SCmod:C with GOL in the iv and v positions and SCmod:D with GOL in the i, did not present negative frequencies in the vibrational spectra (not shown), we concluded that these models represent configurations closer to the global minimum energy without mechanical instabilities. Thus, we argued that the lowest energy SCmod:C with GOL in the v (labelled as SCmod:C(v) herein) will be the suitable model to mimic SC from the energetic and mechanical point of view.
	Figure 3 presents the FTIR experimental data for SC tissue in the fingerprint region (800 - 1800 cm$^{-1}$). The peptidic bond Amide vibrations (1500 - 1700 cm$^{-1}$) dominates the spectra as expected for a sample with high protein content. The solid line is the simulated spectra using the SCmod:C(v) model. The agreement between experimental and simulation is excellent below 1500 cm$^{-1}$. For Amides region, the general spectral vibrations are present; however, important contributions arising from confined water have not been not taken into account in SCmod, which explains the observed difference.
	
	\begin{figure}[th!]
\centering
	\includegraphics[width=8.0cm]{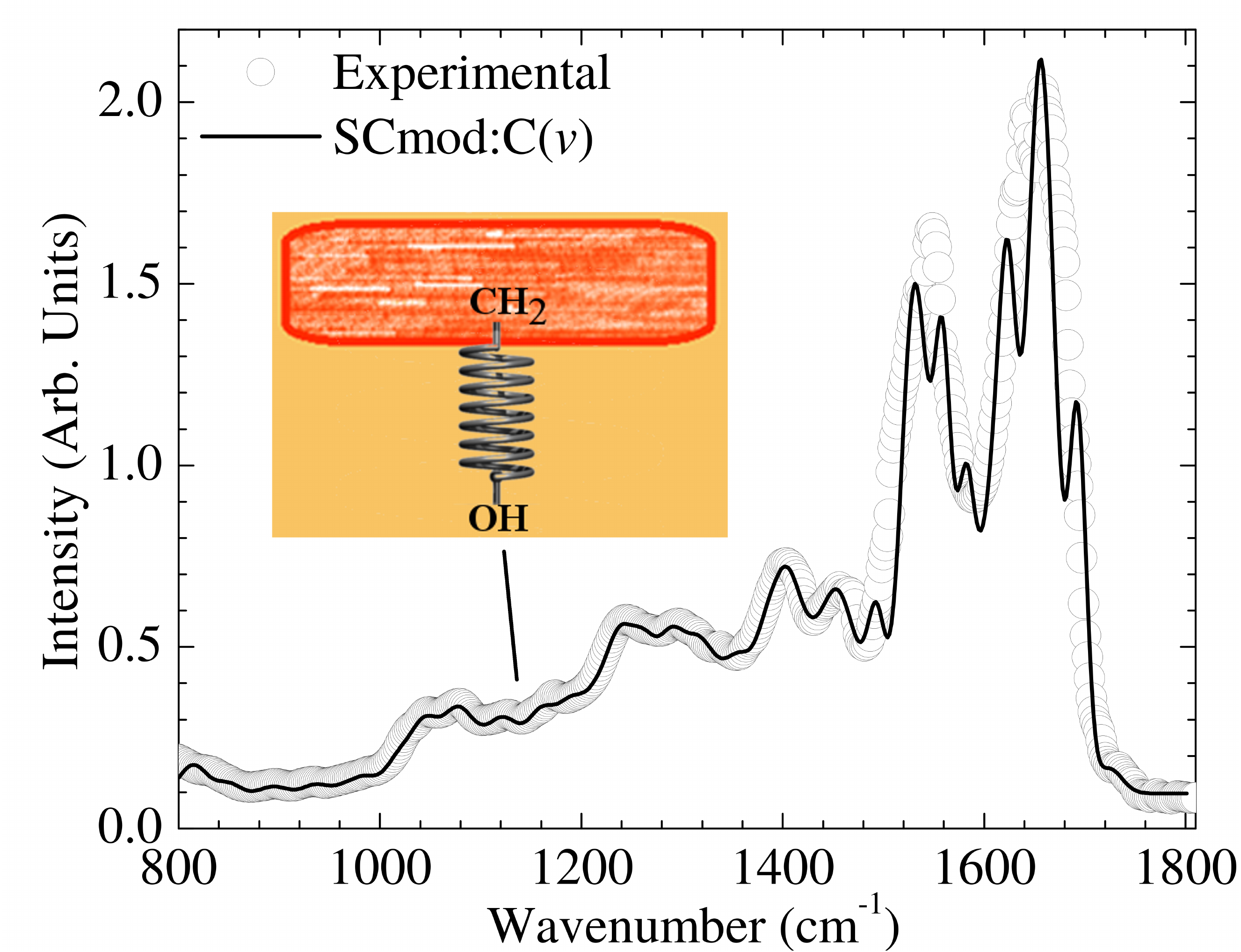}
	\caption{Experimental FTIR spectra for stratum corneum  in the fingerprint region. The solid line is the simulated spectra considering the vibrational calculations for SCmod:C with GOL in the v position (Scmod:C(v)). The inset shows the “mortar”-“lipid” interaction represented by an spring coupling  hydroxil from GOL and methylene from proteic unit cell. The corresponding vibrations will appear in 1160 cm$^{-1}$ frequency according Table 1.}\label{fig3}
\end{figure}

Important pieces of information arise from the vibrational assignment indicated on Table I. This table contains some representative bands (Amides II and IV, some side chains and skeletal characteristics vibrations) since the full assignment would appear exhaustive smearing out the main features captured by the model. Those bands related to the GOL vibrations are indicated as red. The 1120 cm$^{-1}$ band is the hydroxyl stretching methylene wagging coupled vibration of GOL. The next one, at 1160 cm$^{-1}$, has similar symmetry. It is a coupled vibration between hydroxyl group of GOL and NMF methylene. Both bands are of special interest since they would be useful to probe the lipid-amino acid coupling, a parameter of special relevance for studies in Dermatology, Cosmetics, and Biomedical Engineering. For example, it would be used to evaluate the interactions of substances applied in the skin with the proposes to contribute to organize the structure or improve the broken/damaged structures, chance the moisture level, among others. In particular, spectral changes were identified as arising from alterations in the concentration of the major constituents of NMF, important in maintaining SC hydration\cite{Xu2016,Choe2017}.

\begin{table*}[tbh!]
	\caption{\label{assign} Vibrational modes assignment for some FTIR bands observed in stratum corneum tissue (Fig. 3) based on SCmod:C(v) model . Bands indicated in red are of special interest since they probe the brick-mortar interaction.}
\begin{tabular}{|c|c|l|}
\hline 
Experimental & SCmod & Assignment of some bands \\ 
\hline 
1040 & 1040 & (CH$_2$+CH$_3$) asymmetric torsion of proline ring and cage \\ 
\hline 
\textcolor{red}{1120} & \textcolor{red}{1118} &\textcolor{red}{ [OH stretching+CH$_2$ wagging]$^{GOL}$ } \\ 
\hline 
\textcolor{red}{1160} & \textcolor{red}{1160} &\textcolor{red}{ [OH stretching+CH$_2$]$^{GOL}$+CH$_2$ wagging} \\ 
\hline 
1240 & 1238 & [COH-CH$^{\alpha}$-CH$^{\delta_2}$] proline ring symmetric torsion \\ 
\hline 
1400 & 1403 & CH$_2$ twisting ring proline \\ 
\hline 
1500 & 1496 & C-CH$_3$ symmetrical bend (umbrella mode) \\ 
\hline 
1560 & 1561 & Amide II \\ 
\hline 
1580 & 1581 & Amide IV \\ 
\hline 
1620 & 1619 & Amide II+(CH$_2$+CH$_3$) asymmetric torsion (ring and cage) \\ 
\hline 
1650 & 1653 & Amide II+(CH$_2$+CH$_3$) asymmetric torsion (ring and cage) \\ 
\hline 
1700 & 1694 & Amide II+CH$_2$ twisting \\ 
\hline 
\end{tabular} 
\end{table*}

It is well known that thickness and roughness of SC changes with gender, age and anatomical site\cite{Maiti2016}. Thus, determination of mechanical parameters of skin correlated to its structure or biochemical composition is of direct interest\cite{Maiti2016}. The results of our calculations can also be used, e.g., to evaluate mechanical properties of SC. We will present an estimate of the shear modulus (S) of the SC. Considering the corneocyte-lipid interaction as going from NMF amino acids – mortar lipids modeled as an harmonic potential, the perturbation frequency would be approximated by as natural frequency of a spring-mass-like oscillator\cite{French2003} 
	
\begin{equation}
\nu=\frac{1}{2\pi}\sqrt{\frac{S}{l_0}I}
\end{equation}
where $l_0$  and $I$ 	are the corneocyte depth and inertia moment, respectively. Using $\nu=1,160$ cm$^{-1}$ , and $l_0$  and $I$ from ref. \cite{1983} we found $S=0.200$ MPa in good agreement with dry SC indentation measurements from ref. \cite{Leyva-Mendivil2015}. The actual skin mechanical measurements protocols evolves indentations tests, rheometry, and others tribologycal techniques (see, e.g.\cite{Lewis2015,Chen2015,Bostan2016}) which are unespecific to chemical composition and unresolved microscopically.

	As other possible applications of the SCmod proposed model we can cite i) drug delivery studies where previsions about the outcomes of experiments designed to tracking specific molecules crossing SC (e.g., single molecule absorption spectroscopy, vibrational spectroscopy) could be simulated with high accuracy; ii) mechanical properties (like Young's modulus, shear modulus, Poisson's ratio, mechanical toughness) previsions of SC as function of moisture content variations or applied load; iii) study of corneocyte interactions by including more layers into the model. It is useful to understand physical/biochemical properties of tissues since they enable study the molecular interactions which would be present; to quantify the relative strength of each interaction; to perform dynamical simulations; and to predict properties such as electronic structure, transport, mechanical, with high degree of accuracy\cite{sato}.

\section{Conclusions}

We conclude that our proposed SCmod minimalist model for single dry SC is able to capture the main experimental trends of SC as probed by FTIR spectroscopy performed on animal model of SC. In the specific case considered the calculated vibrational spectra enabled assign the main vibrations active observed in SC and reveal specific vibrational probes to the “brick”-”mortar” interactions. It is important to stress that more pieces of information could be explored and analyzed using the large amount of data obtained. A large amount of modes were not described for shake of simplicity. Finally, the high-wavenumber region (up to 4000 cm$^{-1}$) was not considered in the present work because important modes from confined water which are active in this spectral window are silent in our version of dry Scmod.

Research in dermatology, cosmetology, and biomedical engineering in specific fields of drug delivery and/or mechanical properties of skin can be performed taking advantage of chemical accuracy and molecular resolution of our proposed model quantum model. It is important to mention that SCmod could be improved by introducing specific characteristics under study and other physical, chemical and biological properties beyond of the selected examples included in the present paper could be calculated.

\textbf{Acknowledgements.} The authors would like to thank the Brazilian agencies Conselho Nacional de Desenvolvimento Científico e Tecnológico (CNPq - 311146/2015-5) and Fundação de Amparo à Pesquisa do Estado de São Paulo (FAPESP - 2011/19924-2) for the financial support. The authors would also thank the computational resources provided by Centro Nacional de Processamento de Alto Desempenho em São Paulo (CENAPAD-UNICAMP) and Sistema de Computação Petaflópica (Tier 0) (Santos Dumont-LNCC) under  Sistema Nacional de Processamento de Alto Desempenho (SINAPAD) of the  Ministério da Ciência, Tecnologia e Inovação (MCTI).\\

\section{References}

\end{document}